\documentclass[12pt,epsbox]{article}
\usepackage[dvips]{graphicx}
\usepackage{amsmath,amssymb}

\baselineskip=\normalbaselineskip
\renewcommand{\baselinestretch}{1.4}
\newcommand{\resection}[1]
 {\setcounter{equation}{0}\section{\large{#1}}}

\newcommand{\be}{\begin{equation}}
\newcommand{\ee}{\end{equation}}
\newcommand{\ba}{\begin{eqnarray}}
\newcommand{\ea}{\end{eqnarray}}
\newcommand{\nn}{\nonumber \\}
\newcommand{\qq}{\qquad}
\newcommand{\del}{\partial}
\newcommand{\bra}[1]{\left\langle\,{#1}\,\right|}
\newcommand{\ket}[1]{\left|\,{#1}\,\right\rangle}
\newcommand{\bracket}[2]{
\left\langle\left.\,{#1}\,\right|\,{#2}\,\right\rangle}
\newcommand{\eq}[1]{(\ref{#1})}
\newcommand{\bz}{\overline{z}}
\newcommand{\tb}{\widetilde{b}}
\newcommand{\tc}{\widetilde{c}}
\newcommand{\talpha}{\widetilde{\alpha}}
\newcommand{\bb}{\overline{b}}
\newcommand{\bc}{\overline{c}}
\newcommand{\calO}{{\mathcal{O}}}
\newcommand{\hh}{\widehat{h}}
\newcommand{\hD}{\widehat{D}}
\newcommand{\tL}{\widetilde{L}}

\begin{document}
\setcounter{page}{0}
\begin{flushright}
\parbox{40mm}{%
YITP-03-62\\
KEK-TH-912\\
{\tt hep-th/0309074} \\
September 2003}
\end{flushright}

\vfill

\begin{center}
{\Large{\bf 
Closed String Field Theory with Dynamical D-brane 
}}
\end{center}

\vfill

\renewcommand{\baselinestretch}{1.0}

\begin{center}
{\sc Tsuguhiko Asakawa}
\footnote{E-mail: {\tt asakawa@yukawa.kyoto-u.ac.jp}},  
{\sc Shinpei Kobayashi}
\footnote{E-mail: {\tt shinpei@phys.h.kyoto-u.ac.jp}} and 
{\sc So Matsuura}
\footnote{E-mail: {\tt matsuso@post.kek.jp}} 

~\\

$^1${\sl Yukawa Institute for Theoretical Physics, 
      Kyoto University, Kyoto 606-8502, Japan } \\
$^2${\sl Graduate School of Human and Environment Studies, Kyoto
University, Kyoto  606-8501, Japan} \\
$^3${\sl High Energy Accelerator Research Organization (KEK), 
Tsukuba, Ibaraki, 305-0801, Japan}

\end{center}

\vfill

\begin{center}
{\bf abstract}
\end{center}

\begin{quote}

\small{%
We consider a closed string field theory with an arbitrary matter 
current as a source of the closed string field. 
We find that the source must satisfy a constraint equation 
as a consequence of the BRST invariance of the theory.
We see that it corresponds to the covariant conservation law 
for the matter current, 
and the equation of motion together with this constraint 
equation determines 
the classical behavior of both the closed string field and the matter. 
We then consider the boundary state (D-brane) 
as an example of a source. 
We see that the ordinary boundary state 
cannot be a source of the closed string field when the string coupling 
$g$ turns on.
By perturbative expansion, we derive a recursion relation which 
represents the bulk backreaction and the D-brane recoil. 
We also make a comment on the rolling tachyon boundary state.
}
\end{quote}
\vfill

\renewcommand{\baselinestretch}{1.4}

\renewcommand{\thefootnote}{\arabic{footnote}}
\setcounter{footnote}{0}
\addtocounter{page}{1}

\resection{Introduction}

Throughout the recent studies of the (super)string theories, 
we have obtained deep insights into the non-perturbative 
or the off-shell structure of the string theories. 
Especially, 
D-branes have played extremely important roles. 
For example, the studies of BPS saturated D-brane systems 
made us discover the U-dualities between superstrings. 
The AdS/CFT correspondence \cite{ads-cft} and  
the holographic renormalization group \cite{FGPW,dVV} 
(for recent review, see Ref.\ \cite{FMS})  
are also important examples that were found 
by studying D-brane systems. 
In addition to these non-perturbative properties, 
the understanding of the off-shell structure of string theories 
has greatly progressed 
by studying D-branes from the viewpoint of open string theory. 
For example, it was conjectured that 
unstable D-brane systems decay into the vacuum or lower dimensional
D-branes through the tachyon condensation \cite{Sen}, 
and analysis using various methods supports 
the correctness of this conjecture 
(see e.g. \cite{Taylor,BSFT}). 
The rolling tachyon solution was also proposed, 
which is a time dependent background representing 
the rolling down of the open 
string tachyon field towards the bottom of its potential 
\cite{rolling}. 

Another important feature of D-brane is that it is thought to be 
a soliton of closed string theory. 
This is well understood by expressing the D-brane 
as the boundary state \cite{CLNY}\cite{PC}. 
D-brane is originally defined as an object on which 
open strings can attach their end points. 
The corresponding boundary condition is determined 
so that it does not break 
the conformal symmetry of the world-sheet with the disk topology
(the boundary CFT),  
and this symmetry enable us to transform the boundary condition for 
open strings into that for closed stings.   
The obtained state $\ket{B}$ is the boundary state which satisfies 
the boundary condition in terms of closed strings.
Then, it can be viewed as a source in closed string theory
\cite{CLNY}\cite{PC}. Namely,  
adding a boundary to the world-sheet is equivalent to adding  
a boundary state to the equation of motion for the 
closed string field as
\begin{equation}
 Q\ket{\Phi} = -\ket{B},
\label{free-eom} 
\end{equation}
where $Q$ is the BRST charge of the closed string.
The nilpotency $Q^2=0$ implies that the admissible boundary states  
are characterized by the condition,   
\begin{equation}
 Q\ket{B} = 0, 
\end{equation}
that is, BRST invariant boundary states give conformal backgrounds.

Open string dynamics on the D-brane 
can also be expressed by inserting an appropriate boundary 
interaction in the boundary state 
and they correctly describes the tachyon condensation
which is mentioned above (see e.g. \cite{AST}). 
The rolling tachyon background can be also described in the same way 
\cite{rolling}.
Through the study of this rolling tachyon boundary state, 
it is found that the final state
of this decay is not the closed string vacuum 
but a state with finite energy density and no pressure,  
which is called the tachyon matter \cite{t-matter}. 
However, in spite of many studies on the rolling tachyon 
and the tachyon matter \cite{OY,OS,LLM}, 
the relation between the closed string emission from 
the decaying D-brane and the tachyon condensation is not clear yet. 
This would be because the 
effect of closed string interactions are not taken into account
in \eq{free-eom}.

One of the main purpose of this article is to give a general formalism 
to deal with D-branes in a closed string field theory (closed SFT).%
\footnote{
For another approach, see Ref.\ \cite{HH}\cite{Zwie}. 
}  
In other words, we will see what happens to \eq{free-eom} 
 by turning on the closed string coupling $g$.
In this article, 
we regard the boundary state as a matter current 
which couples to the closed string field. 
We first give a general formalism to determine 
the classical behavior of the closed string field 
when there is an arbitrary matter current 
which couples to the closed string field. 
Starting with adding a source term to the action of a closed SFT, 
we find a constraint equation that the source must satisfy and we see 
that the constraint equation plays an important role in this formalism.  
Although we adopt HIKKO's closed SFT \cite{hikko} as an example of 
a closed SFT because of its simplicity, 
we emphasize here that 
our argument does not depend on the detail of the theory 
but applicable to any kind of closed SFTs
that is consistent at least at the tree level 
in the sense of BRST invariance, 
because our argument relies only on the BRST invariance 
of the theory in the tree level. 
One of our most interesting results is that 
the ordinary boundary state does not satisfy the constraint equation 
but must be modified so that it can be a consistent source 
of the closed string.  
We see that it is quite natural to expect that 
the modification is caused by open string excitations on the D-brane, 
which give dynamical degrees of freedom to the source.

The organization of this paper is the following. 
In \S 2, we give a brief review of HIKKO's closed SFT 
in order to confirm the notation that we use in this article. 
We explain the BRST and the gauge symmetry of 
the closed SFT in detail. 
In \S 3, we add a source term to the SFT action 
and derive the constraint equation mentioned in the last paragraph. 
We show that the equation follows from the nilpotency of 
the BRST transformation of the SFT. 
Applying the analysis to a boundary state, 
we show that the ordinary boundary state cannot be 
a source of the closed string field 
unless the closed string coupling $g$ vanishes, 
and it must be modified by the interaction of the 
closed string field with the boundary 
so as to satisfy the constraint equation. 
We claim that the modification occurs as a consequence of 
open string excitations on the D-brane. 
We also make a comment of the rolling tachyon boundary state 
\cite{rolling, t-matter}. 
The section 4 is devoted to the conclusion and discussions. 
In the appendix A, we explain the construction of the 
$*$-product of HIKKO's SFT in detail. 
In the appendix B, we show explicitly that the free closed SFT 
actually reproduces the quadratic terms of the gravity theory 
if we restrict the closed string field up to the massless level. 
We also show that the gauge transformation of the free closed SFT 
correctly reproduces that for the fields in the gravity theory.

\resection{Review of Closed String Field Theory}

In this section, we briefly review a bosonic closed string field theory, 
that is discussed in Ref.~\cite{hikko} (HIKKO's closed SFT) 
in order to confirm our notations. 
We mainly follow the convention in Refs.\ \cite{KT, AKT}, 
where the familiar conformal field theory (CFT) 
language is used to describe string field theories. 
We explain the ghost zero-mode structure of the string field and 
the BRST invariance of the action in detail, 
which are frequently used in later sections.  
We note that the discussion in the following section does not 
depend on the detail of HIKKO's theory but only use 
the BRST invariance of the closed SFT (see below). 
The reason we use HIKKO's closed SFT is only its simplicity. 

Let us start with fixing the convention of the CFT that defines 
the bosonic string theory in the flat $26$ dimensional space-time.  
The elementary fields are  
the 26 scalar fields $X^{\mu}(z,\bz)$, 
the holomorphic ghost fields $b(z)$ and $c(z)$, 
and the antiholomorphic ghost fields $\bb(\bz)$ and $\bc(\bz)$. 
If we set $\alpha'=2$, the mode expansions are \cite{Pol}
\begin{align}
 \del X^{\mu}(z) &= -i 
 \sum_{n=-\infty}^{\infty}\frac{\alpha_n^{\mu}}{z^{n+1}}\,, \nn
  b(z) &= \sum_{n=-\infty}^{\infty}\frac{b_n}{z^{n+2}}\,, \\
  c(z) &= \sum_{n=-\infty}^{\infty}\frac{c_n}{z^{n-1}}\,,  \nonumber
\end{align}
and the antiholomorphic fields are similarly expanded 
into the oscillators $\{\talpha_n^\mu, \tb_n, \tc_n \}$. 
These oscillators satisfy the algebra, 
\begin{align}
 \left[\alpha_m^{\mu},\alpha_n^{\nu}\right] 
  &= \left[\talpha_m^{\mu},\talpha_n^{\nu}\right] 
  = m\,\eta^{\mu\nu}\delta_{m+n\,,\,0}\,, \\
 \left\{b_m, c_n\right\} 
  &= \left\{\tb_m, \tc_n\right\} 
  = \delta_{m+n\,,\,0}\,. 
\end{align}
In this article, we adopt the following 
notation for the ghost zero-modes;  
\begin{alignat}{2}
  c_0^+ &\equiv {1\over 2}\left(c_0 + \tc_0\right)\,, &\qq 
  c_0^- &\equiv c_0 - \tc_0\,, \nn 
  b_0^+ &\equiv b_0 + \tb_0\,, &
  b_0^- &\equiv {1\over 2}\bigl(b_0 - \tb_0\bigr)\,,   
\end{alignat}
which satisfy 
\begin{equation}
 \left\{b_0^{\pm},c_0^{\pm}\right\}=1 \,.
\end{equation} 
An arbitrary state in the Hilbert space of this CFT is 
obtained by acting some numbers of the oscillators on the 
$SL(2,\mathbf{C})$-vacuum $\ket{0}$ which satisfies  
\begin{equation}
 \alpha_n^{\mu}\ket{0}=0 \quad (n \ge 1), \qq
  b_n\ket{0}=0 \quad (n \ge -1), \qq
  c_n\ket{0}=0 \quad (n \ge 2). 
\end{equation} 
As usual, we assign the ghost number 1 for $c(z)$ and $\bc(\bz)$ 
and $-1$ for $b(z)$ and $\bb(\bz)$. 
We also set the ghost number for the $SL(2,\mathbf{C})$-vacuum 
to be zero. 
Then, any physical states in the bosonic string theory 
(e.g., the tachyon state $c_1\tc_1\ket{k}$) 
have ghost number $2$. 
Because of the ghost number anomaly on $S^2$, any non-zero matrix
element should have ghost number $6$. 
Then we take a convention,%
\footnote{
The absence of an $i$ in the right hand side is compensated 
by the following unusual definition of the Hermitian conjugate
\cite{Zwiebach},
\begin{equation}
 \left(\bracket{\Phi_{\rm hc}}{\Psi}\right)^{\dagger}
 = - \bracket{\Psi_{\rm hc}}{\Phi}, 
\end{equation}
where $\bra{\Phi_{\rm hc}}$ expresses the Hermitian conjugate of 
the state $\ket{\Phi}$.}
\begin{equation}
 \bra{k'}c_{-1}\tc_{-1}c_0^-c_0^+c_1\tc_1\ket{k} 
  = (2\pi)^{26}\delta^{26}(k-k').  
  \label{overlap}
\end{equation}

Now we define a closed string field 
using the language of the CFT described above. 
Roughly speaking, an arbitrary closed string field is a vector 
in the Hilbert space of the above CFT, expressed as 
a linear superposition of the basis states  
with coefficients as target space fields. 
Additionally, the closed string field must satisfy the following 
two constraints \cite{Zwiebach}, that is, 
the level matching condition, 
\begin{equation}
 L_0^-\ket{\Phi} \equiv {1\over2}\left(L_0-\tL_0\right)\ket{\Phi}= 0\,, 
  \label{rotational-inv}
\end{equation} 
and the reality condition, 
\begin{equation}
 \bra{\Phi} = \bra{\Phi_{\rm hc}}\,,
  \label{reality}
\end{equation}
where 
$L_0\,(\tL_0)$ is the zero-mode of the (anti)holomorphic 
Virasoro generators of the CFT 
and $\bra{\Phi}$ and $\bra{\Phi_{\rm hc}}$ are 
the BPZ conjugate%
\footnote{
The BPZ conjugate is defined via the conformal mapping $I(z)=1/z$. 
The BPZ conjugate of the state $\ket{\calO}\equiv {\calO(z=0)}\ket{0}$ 
is $\bra{\calO}\equiv \bra{0} I[\calO](z=0)$. }
and the Hermitian conjugate of $\ket{\Phi}$, 
respectively.

The closed string field can be decomposed into four sectors 
corresponding to the degeneracy of the closed string vacua 
due to the presence of the ghost zero-modes as 
\begin{equation}
 \ket{\Phi} = c_0^- \bigl(\ket{\phi} + c_0^+ \ket{\psi} \bigr)
  + \bigl(\ket{\chi} + c_0^+ \ket{\eta}\bigr). 
\end{equation}
However, in writing down the action of a string field theory with 
$\ket\Phi$, 
we need only two of these sectors, and thus we impose another condition,%
\footnote{Our notation is different from that of \cite{Zwiebach} 
where physical states are in the $\ket{\chi}$ sector, i.e. 
$b_0^-\ket{\Phi}=0$ is imposed. As a consequence of this, 
various definitions below are different. However, we can easily change
the convention, 
and the consequence of this paper is not affected by
the choice of the convention.}
\begin{equation}
 c_0^-\ket{\Phi} = 0, 
\end{equation}
that is, 
\begin{equation}
 \ket{\Phi} = c_0^- \ket{\phi} + c_0^-c_0^+ \ket{\psi} \,. 
 \label{decomposition}
\end{equation} 
We assume that 
the physical target-space fields (dynamical variables) are 
in the sector $\ket{\phi}$. 
Then the ghost number of $\ket{\phi}$ turns out to be $2$ 
and it becomes Grassmann even. 
As a result, the ghost number of $\ket{\Phi}$ is $3$, 
while that of the sector $\ket{\psi}$ is $1$, 
which are both Grassmann odd. 
As we will see below, the target-space fields in the sector
$\ket{\psi}$ are auxiliary fields. 
The fact that a string field has two sectors plays 
important role in the next section.
In the following, 
in addition to the bracket notation above, 
we also denote a string
field as a functional $\Phi$ with CFT fields as the coordinate. 
We freely use both expressions below.

Next, we give the action of HIKKO's closed string field
theory and describe its symmetries. The action is written as 
\begin{equation}
 S = {1\over 2}\Phi\cdot Q\Phi + {g\over 3}\Phi\cdot\Phi*\Phi \,, 
 \label{action}
\end{equation}
Here $g$ is the closed string coupling constant. 
$Q$ is the (total) BRST charge in the flat background and is
nilpotent $Q^2=0$.
It is decomposed by the ghost zero-modes as
\begin{equation}
 Q = c_0^+ L_0^+ + b_0^+ M^+ + Q' + \cdots \,,
  \label{BRST-charge}
\end{equation}
with
\begin{align}
 L_0^+ & \equiv L_0+\tL_0\,,\\ 
 M^+ &\equiv -\sum_{n=1}^{\infty} n\left(c_{-n}c_n +
 \tc_{-n}\tc_n \right) \,, \\
 Q' &\equiv \sum_{n\ne 0}
 \left(c_{-n}L_n^{(m)}+\tc_{-n}\widetilde{L}_n^{(m)}\right) \,. 
 \label{Qdef}
\end{align}
Here, we have denoted the (anti)holomorphic Virasoro 
generators of the matter CFT as 
$\{L_n^{(m)}\,(\widetilde{L}_n^{(m)})\,|\, n\in{\mathbf Z}\}$.%
\footnote{
If we write the Virasoro generators of the ghost CFT as 
$\{L_n^{(g)}\,(\widetilde{L}_n^{(g)})\,|\, n\in{\mathbf Z}\}$, 
the total Virasoro generators are expressed as 
$L_n=L_n^{(m)}+L_n^{(g)}-\delta_{n,0}$.}
The ``$\cdots$'' in \eq{BRST-charge} contains terms with $b_0^-$ and $c_0^-$, 
which have no effect to the action \eq{action}. 
The inner product $\cdot$ of string 1 ($\ket{\Phi}_1$) and string 2 
($\ket{\Psi}_2$) is defined as
\begin{align}
 \Psi \cdot \Phi 
 &\equiv \bra{R(1,2)}b_0^{(2)-}\ket{\Psi}_2\ket{\Phi}_1 \nn
 &= \bra{\Psi}b_0^-\ket{\Phi}, 
 \label{inner}
\end{align}
where the superscript of $b_0$ means that the oscillator $b_0$ belongs to 
string 2, and $\bra{R(1,2)}$ is the reflector that maps 
an arbitrary state 
$\ket{{\mathcal{O}}}$ to its BPZ conjugate  $\bra{{\mathcal{O}}}$, 
\begin{equation}
 \bra{R(1,2)}\ket{{\mathcal{O}}}_2 = {}_1\bra{{\mathcal{O}}}.
\end{equation}
The $*$-product is defined as a mapping from two string fields 
to one string field, which is written as  
\begin{equation}
\ket{\Phi}*\ket{\Psi} \equiv \ket{\Phi*\Psi}\,.
\end{equation} 
The more precise definition of the $*$-product in HIKKO's SFT is 
summarized briefly in Appendix \ref{star-product}. 
However, the details of the $*$-product is not necessary 
in this article. 
We only need below is the following properties proved in  
Ref.\ \cite{hikko}, 
\begin{align}
 (1) \quad & \Phi\cdot\Psi 
     = (-1)^{\left|\Phi\right|\left|\Psi\right|}\,\Psi\cdot\Phi \,,  \\
 (2) \quad & Q\Phi\cdot\Psi 
      = -(-1)^{\left|\Phi\right|}\Phi\cdot Q\Psi \,, \\
 (3) \quad & {\Phi*\Psi} = -(-1)^{\left|\Phi\right|\left|\Psi\right|}\,
                               {\Psi*\Phi}\,, \\ 
 (4) \quad & Q\left(\Phi*\Psi\right) = Q\Phi*\Psi 
     + (-1)^{\left|\Phi\right|}\Phi * Q\Psi \\
 (5) \quad & (-1)^{\left|\Phi\right|\left|\Lambda\right|}
             \left(\Phi*\Psi\right)*\Lambda 
            +(-1)^{\left|\Psi\right|\left|\Phi\right|}\, 
             \left(\Psi*\Lambda\right)*\Phi
            +(-1)^{\left|\Lambda\right|\left|\Psi\right|}\, 
             \left(\Lambda*\Phi\right)*\Psi = 0 \,, \\
 (6) \quad & \Lambda\cdot(\Phi*\Psi) 
           =(-1)^{\left|\Lambda\right|\,\left(\left|\Phi\right|
                 +\left|\Psi\right|\right)}\, 
            \Psi\cdot\left(\Phi*\Lambda\right) 
           =(-1)^{\left|\Phi\right|\,\left(\left|\Psi\right|
                 +\left|\Lambda\right|\right)}\, 
            \Phi\cdot\left(\Psi*\Lambda\right) \,, 
\end{align}
where $\left|\Phi\right|$ represents the Grassmann parity of 
the closed string field $\ket{\Phi}$. 

It is useful to see that, in the action \eq{action}, 
dynamical fields in the target-space are 
actually in the physical sector $\ket{\phi}$. 
Substituting \eq{decomposition} into 
the free part of the action \eq{action}, 
we obtain 
\begin{align}
 S_0 &= {1\over2} \Phi\cdot Q\Phi  \nn
     &= {1\over2} \bra{\phi}c_0^-c_0^+L_0^+\ket{\phi} 
        -{1\over2} \bra{\psi}c_0^-c_0^+M^+\ket{\psi}
        -\bra{\psi}c_0^-c_0^+Q'\ket{\phi}. 
 \label{kinetic}
\end{align} 
Recalling that only $L_0^+$ contains a term quadratic in momentum 
(or space-time derivative $\sim \del^2$),
we see that only fields in $\ket{\phi}$ have kinetic terms, 
and thus, dynamical fields are surely in $\ket{\phi}$. 
On the other hand, since the second and the third terms of \eq{kinetic} 
have terms at most linear in the momentum, 
it can be understood that 
target-space fields in $\ket{\psi}$ are auxiliary fields.  
More explicitly, if we restrict the string field 
up to the massless level, 
we can show that the free action 
\eq{kinetic} reproduces the quadratic part of 
the low energy effective action of the 
bosonic string theory \cite{GS}. 
We perform it explicitly in the Appendix \ref{freeSFT}.

We next discuss the BRST and gauge symmetry of the action \eq{action}. 
They are governed by the general structure of the Batalin-Vilkovisky
formalism \cite{BV}. First we define the
BRST transformation \cite{BV} (see also \cite{AKT}),%
\footnote{It is also called as pre-BRST transformation or master
transformation.} 
\begin{equation}
\delta_{\rm B} b_0^-{\Phi} \equiv \frac{\delta}{\delta {\Phi}} S,   
\end{equation}
then the BRST transformation of $\Phi$ in HIKKO's closed SFT turns out to be  
\begin{equation}
 \delta_{\rm B} \Phi =  Q\Phi + g\Phi*\Phi. 
  \label{BRST}
\end{equation} 
The most important property of the BRST transformation is 
its nilpotency, and it is a direct consequence of 
the properties (3), (4) and (5) above, 
\begin{align}
 \delta_{\rm B}^2 \Phi &= \delta_{\rm B}\left(
 Q\Phi + g\Phi*\Phi
 \right) \nn 
 &= -Q\left(Q\Phi+g\Phi*\Phi\right)
    +2g\left(Q\Phi+g\Phi*\Phi\right)*\Phi \nn 
 &= -Q^2\Phi + g\Bigl[-Q\left(\Phi*\Phi\right)+2Q\Phi*\Phi\Bigr] 
    + 2g^2\Bigl[\left(\Phi*\Phi\right)*\Phi\Bigr] \nn 
 &= 0 \,.
  \label{nilpotent}
\end{align}
Properties (1)$\sim$(6) guarantee that the action $S$ is the solution
to so called (classical) BV master equation. 
Moreover, the nilpotency of the BRST transformation \eq{nilpotent} 
is equivalent to the BRST invariance of the action \eq{action},
\begin{equation}
 \delta_{\rm B}S = 0.
\end{equation} 
This means that it is not necessary to add more interaction terms 
to the action \eq{action} at least at the tree-level. 
Note that any other SFT with the BV structure defines its own BRST
transformation and has the same property.

The nilpotency of the BRST transformation also guarantees 
the gauge invariance of the action under the 
gauge transformation 
\begin{equation}
 \delta_{\Lambda}\Phi \equiv Q\Lambda + 2g \Phi*\Lambda.
  \label{gauge-trans}
\end{equation}
Here $\Lambda$ is a gauge parameter, 
which is a closed string field with the ghost number two. 
To clarify the structure of the gauge transformation, 
let us decompose the first term of (\ref{gauge-trans}) 
in terms of $\ket{\phi}$ and $\ket{\psi}$. 
To this end, we expand $\ket{\Lambda}$ as 
\begin{equation}
 \ket{\Lambda} = c_0^-\ket{\lambda_1} + c_0^-c_0^+\ket{\lambda_2}\,. 
\end{equation} 
Then we see that 
the gauge transformations for $\ket{\phi}$ and $\ket{\psi}$ become 
\begin{align}
  \label{gauge-phys}
 \delta \ket{\phi} &= -Q'\ket{\lambda_1} - M^+ \ket{\lambda_2}\,, \\ 
 \delta \ket{\psi} &= Q'\ket{\lambda_2} - L_0^+ \ket{\lambda_1}\,. 
\end{align}
Recalling that dynamical target-space fields are in $\ket{\phi}$ 
and $Q'$ contains the target-space differential in the first
order, it turns out that 
the gauge parameters of the dynamical fields are in $\ket{\lambda_1}$. 
Moreover, from the second term of \eq{gauge-phys}, 
we see that some of the target-space fields 
in the physical sector $\ket{\phi}$ can be gauged away 
using the degree of freedom of $\ket{\lambda_2}$.%
\footnote{
An example of such a field is $S(k)$ in the decomposition, 
\[
\ket{\phi} = \int \frac{d^{26}k}{(2\pi)^{26}} \left[\cdots 
- {1\over\sqrt{2}}S(k)\left(c_{-1}c_1+\tc_{-1}\tc_1\right)
+ \cdots\right]\ket{k}. 
\]
For detail, see the Appendix \ref{freeSFT}. }
The gauge transformation of the low energy fields are also
discussed explicitly in the Appendix \ref{freeSFT}.

\resection{Source Term in Closed String Field Theory}

In this section, we discuss the general structure of 
HIKKO's closed SFT with a source term. 
We first consider a closed SFT action with a source term and 
derive two equations that the closed string field and the source 
should satisfy classically. 
After that, we consider a boundary state as a source 
of the closed string field. 
We also make some comment on the rolling tachyon boundary state
\cite{rolling} from the view point of the closed SFT.  


\subsection{Constraint to A Source of Closed String Field}

We start with the action of HIKKO's closed SFT (\ref{action}), 
which is invariant under the BRST transformation (\ref{BRST}) 
and gauge transformation (\ref{gauge-trans}). 
We then add to it a source term as
\begin{equation}
 S = {1\over 2}\Phi\cdot Q\Phi + {g\over 3}\Phi\cdot\Phi*\Phi 
  + \Phi \cdot J\,.
  \label{action-source}
\end{equation}
Here $J$ is considered to be some (yet unknown) matter current. 
In order that $J$ correctly couples to the string field, 
it must be a state in the same Hilbert space 
as string fields live in.
Therefore, $\ket{J}$ should satisfy the level matching condition 
$L_0^- \ket{J}=0$ 
and the reality condition 
$\bra{J}=\bra{J_{\rm hc}}$
as \eq{rotational-inv} and \eq{reality}.
As for the closed string field, we expand $J$ by 
the ghost zero modes as 
\begin{equation}
 \ket{J}=c_0^-\ket{j_\psi}+c_0^- c_0^+\ket{j_\phi}. 
  \label{J}
\end{equation}
Using (\ref{decomposition}), 
we see that $\ket{j_\psi}$ and $\ket{j_\phi}$ couple to the sectors  
$\ket{\psi}$ and $\ket{\phi}$, respectively; 
\begin{eqnarray}
 \Phi\cdot J &=&\bra{\Phi} b_0^- \ket{J} \nonumber \\
 &=& \bra{\psi}c_0^- c_0^+\ket{j_\psi}-\bra{\phi}c_0^- c_0^+\ket{j_\phi}.
  \label{source-component}
\end{eqnarray} 
Recalling that the total ghost number should be $6$, 
we see that 
$j_\psi$ and $j_\phi$ 
must carry ghost number $3$ and $2$, respectively.
This means that $J$ has ghost number $4$.

Applying the variational principle to the action \eq{action-source}, 
the equation of motion of this system is obtained:
\begin{equation}
 Q{\Phi} + g {\Phi*\Phi} + {J} = 0.  
  \label{eom-source}
\end{equation} 
Since the bulk part of this equation of motion
transforms covariantly under the gauge
transformation (\ref{gauge-trans}), the current should also 
transform as 
\begin{equation}
 \delta_{\Lambda}{J} = 2g{J*\Lambda}\,.
 \label{J-gauge}
\end{equation} 
Here, it must be noted that the equation of motion 
\eq{eom-source} is not consistent for arbitrary $J$  
but it must satisfy a consistency condition. 
To find it, let us act the BRST charge 
$Q$ on the left hand side of (\ref{eom-source}), 
\begin{align}
 0 &= Q\bigl(Q{\Phi} + g {\Phi*\Phi} + {J} \bigr) \nn 
   &= Q\left(J+g\Phi*\Phi\right) \nn
   &= QJ+2g \Phi *J \,. 
 \label{constraint}
\end{align}
From the first line to the second line, 
we have used $Q^2=0$, 
and from the second line to the third line, 
we have used the equation of motion \eq{eom-source} 
and the identity, $\left(\Phi*\Phi\right)*\Phi=0$.

To understand what the equation \eq{constraint} means, 
it is useful to consider the (non-abelian) Chern-Simons theory, which
has a formal analogy with our situation. 
Its action with a source is 
\begin{equation}
  S = \int {1\over2} A\wedge dA + {g\over3}A\wedge A \wedge A 
      + A \wedge J\,,  
\end{equation}
where $A$ is some Lie algebra valued $1$-form and matter current $J$
is a $2$-form.
By using the covariant derivative $D \equiv d + gA\wedge$, the equation of motion of this system is given by 
\begin{equation}
 F \equiv DA= -J.
\end{equation}
Using the Bianchi identity $DF=0$, 
it is straightforward from the equation of motion 
to show that the current should be covariantly conserved:
\begin{equation}
 DJ = dJ + g\left(A\wedge J + J\wedge A\right)= 0.
 \label{DJ} 
\end{equation}

This analogy tells us that the BRST transformation 
$\delta_{\rm B}=Q+g\Phi *$ plays the same role of the covariant
derivative  
and the ``Bianchi identity'' corresponds to the nilpotency of the BRST 
transformation $\delta_{\rm B}^2=0$. 
Then, not only the equation of motion \eq{eom-source}, 
we must impose the ``covariant conservation law'' 
\eq{constraint} to $J$ 
as a consequence of the ``Bianchi identity'' \eq{nilpotent}. 
In fact, using the definition of the BRST transformation \eq{BRST} 
and the fact that (\ref{eom-source}) can be written as 
$\delta_{\rm B}\Phi=-J$, we obtain 
\begin{eqnarray}
 0&=& \delta_{\rm B}^2 \Phi \nn 
  &=& -Q\delta_{\rm B} \Phi 
 -2g \Phi *\delta_{\rm B} \Phi \nonumber \\
 &=& QJ+2g\Phi *J. 
\end{eqnarray} 
From this equation, 
although the BRST transformation for the current $J$ has not been defined, 
we can symbolically rewrite this as
\begin{equation}
  \delta_{\rm B} J \equiv QJ+2g\Phi*J = 0,
 \label{conservation}
\end{equation} 
which corresponds to the covariant conservation law for the current 
(\ref{DJ}).
Note that 
the same discussion can also be applied to
any other type of string field theory which has own BRST symmetry, 
since we have only used the equation of motion 
and the nilpotency of the BRST transformation in this derivation. 
In any case, the covariant conservation law takes the form
$\delta_{\rm B}J=0$.


The physical meaning of the equation \eq{conservation} 
can be better understood in the corresponding 
low energy effective theory: 
We can expect that the low energy counterpart of 
the equation \eq{conservation} would be 
the covariant conservation law of the energy-momentum tensor 
in the general relativity,  
\begin{equation}
 \nabla^{\mu}\, T_{\mu\nu} = 0, 
  \label{EM-tensor}
\end{equation} 
where $T_{\mu\nu}$ is the matter energy-momentum tensor. 
One reason which supports it is that 
there is a one-to-one correspondence between each step 
of the derivations of \eq{EM-tensor} and \eq{conservation}.
As is well known, the equation \eq{EM-tensor} can be derived from 
the Einstein equation, 
\begin{equation}
 R_{\mu\nu} - {1\over2}R g_{\mu\nu} = \kappa T_{\mu\nu},   
  \label{Einstein}
\end{equation} 
together with the Bianchi identity, 
\begin{equation}
 \nabla^{\mu}\left(R_{\mu\nu} - {1\over2}R g_{\mu\nu}\right) = 0. 
  \label{Bianchi-grav}
\end{equation} 
On the other hand, as mentioned above,
the equation \eq{conservation} is a consequence of 
the equation of motion \eq{eom-source} 
and the nilpotency of the BRST transformation \eq{nilpotent}. 
Since the closed SFT contains graviton as a massless field, 
it is believed that at low energy \eq{action-source} reduces to 
the Einstein-Hilbert action with some matter source.%
\footnote{We here ignore other massless fields and tachyon, for
simplicity. Note also that the graviton in the SFT and that of
Einstein action is in general related by the non-linear field 
redefinition \cite{GS}.
}
Moreover, the general covariance is a part of the gauge symmetry of
the bulk SFT, which is guaranteed by the nilpotency \eq{nilpotent}. 
Therefore, it is plausible to regard the equation \eq{EM-tensor} 
as the low energy counterpart of the equation \eq{conservation}. 
Another reason supporting this conjecture is 
the direct decomposition of the equation \eq{conservation} 
into the component fields.
To see this,   
let us decompose the physical sector of the closed string field 
as \eq{component-phi} and look only the graviton part. 
Correspondingly, 
the components of the source $J$ which couple to the graviton
through \eq{source-component} are given by
\begin{equation}
 \ket{j_\phi} = \int d^{26}x \left[
 A_{\mu\nu}(x)\left(\alpha_{-1}^{\mu}\talpha_{-1}^{\nu} 
                    +\alpha_{-1}^{\nu}\talpha_{-1}^{\mu}\right)
 +B(x)\left(b_{-1}\tc_{-1}+\tb_{-1}c_{-1}\right)
 +\cdots
 \right]c_1\tc_1\ket{x},
 \label{source-decomp} 
\end{equation}
where $A_{\mu\nu}$ and $B$ are arbitrary functions. 
Their combination 
$T_{\mu\nu}(x)\equiv A_{\mu\nu}(x)+\eta_{\mu\nu}B(x)$ 
is the leading part of the energy-momentum tensor \cite{t-matter}.
Then one can roughly estimate the equation \eq{conservation} as%
\footnote{We ignored corrections comes from other component, 
higher derivative terms and so on. 
We also neglect numerical coefficients in 
front of the second and third term, which are highly dependent on 
the detail on the $*$-product. 
} 
\begin{equation}
    \left( \del T\right)_{\mu}(x) 
  + g\left[
    \left( h \cdot \del T\right)_{\mu}(x)
  + \left( \del h \cdot T\right)_{\mu}(x) 
     \right]=0 \,.  
  \label{h-T}
\end{equation}
On the other hand, if we expand the metric as 
$g_{\mu\nu}=\eta_{\mu\nu}+\kappa h_{\mu\nu}$, 
the equation \eq{EM-tensor} has the same tensor structure
as above at the first order in $h_{\mu\nu}$.

Now we understand that the condition \eq{conservation} is a
generalization of the covariant conservation law \eq{EM-tensor} 
to the SFT, which includes all contribution from massive
fields.  
In the same sense, the second line of \eq{constraint}; 
\begin{equation}
 Q\left(J+g\Phi*\Phi\right)=0\,, 
  \label{total-conservation}
\end{equation}
would be the counterpart of the total energy conservation in gravity 
theory originating both from the matter and from the 
self-gravitating energy; 
\begin{equation} 
 \del^{\mu}\left[\sqrt{-g} 
	    \left(T_{\mu\nu}+t_{\mu\nu}\right)
	   \right]=0 \,, 
 \label{total-grav}
\end{equation}
where $t_{\mu\nu}$ is so called the 
{\em gravitational energy-momentum pseudotensor density}. 
In the gravity theory, the conservation law \eq{total-grav} 
is a direct consequence of the diffeomorphism invariance of 
the total system of the gravity and the matter. 
Correspondingly, the equation \eq{conservation} should be 
a consequence of a gauge symmetry of the SFT. 
Although the SFT action \eq{action-source} is not invariant 
under the gauge transformation \eq{gauge-trans} and \eq{J-gauge}, 
the complete system which includes both the closed string field 
and the matter field 
({\em e.g.} an open-closed SFT) must have a gauge symmetry.  
Once the action of the complete system is given, 
the equation \eq{conservation} would also be required as 
a consequence of the gauge symmetry.

Here, we emphasize that the matter current $J$ 
is considered not to be an external source but a dynamical one. 
Therefore, in solving the equation of motion, 
we should also take into account the covariant conservation law, 
that is, the equation \eq{conservation}. 
Of course, if the full action of the system with 
the closed string and the matter is explicitly given, 
the covariant conservation law of the matter current will be 
automatically satisfied as a consequence of the equation of motion 
of the matter. 
However, since it is not the case now, 
we must solve the equations \eq{eom-source} 
and \eq{conservation} simultaneously.

\subsection{Boundary State as A Source}

From now on, we restrict our attention to boundary states 
and regard them as sources for the closed string field.   
To be more precise, we consider the boundary state $\ket{B_p}$ 
which describes a (bosonic) D$p$-brane, 
extended in $x^\alpha\ (\alpha=0,\cdots,p)$ direction and 
sitting at $x^i=0\ (i=p+1,\cdots,25)$. 
As explained in the introduction, the boundary state is 
obtained by performing the modular transformation for 
the boundary condition of open strings. 
In the above case, we impose the Neumann boundary conditions 
for $X^{\alpha}$ and the Dirichlet boundary conditions 
for $X^i$.
The boundary conditions for the ghost fields
are determined so as the total boundary state is 
BRST- invariant, 
\begin{equation}
Q\ket{B_p}=0.
\label{QB=0} 
\end{equation}
Then, the obtained state is (see, {\em e.g.}, Ref.\ \cite{review})  
\begin{equation}
 \ket{B_p}\equiv \frac{T_p}{2} \delta^{25-p}(x^i)
  \exp\left\{
   \sum_{n=1}^{\infty}\left(
   \frac{-1}{n}\alpha^{\mu}_{-n}S_{\mu\nu}\talpha_{-n}^{\nu}
   +c_{-n}\tb_{-n}+\tc_{-n}b_{-n}
  \right)
 \right\}c_0^+c_1\tc_1\ket{0}\,, 
  \label{b.s}
\end{equation}
where $T_p$ is the tension of the Dp-brane and 
$S_{\mu\nu} \equiv (\eta_{\alpha\beta},-\delta_{ij})$. 
It is a state in the closed string Hilbert 
space with ghost number $3$.
We note that equation \eq{QB=0} consists of two
equations decomposed by the ghost zero-modes. 
In fact, acting the BRST charge \eq{BRST-charge} on \eq{b.s}, 
we see that 
the boundary state satisfies following equations separately:
\begin{equation}
Q'\ket{B_p}=0, \quad b_0^- M^+ \ket{B_p}=0.
\label{MB=0}
\end{equation}


In order to regard the boundary state as a source 
for the physical sector of the closed string field $\ket{\phi}$,
it is first required to multiply $c_0^-$ to \eq{b.s}; 
\begin{equation}
 \ket{J} \equiv c_0^-\ket{B_p}. 
 \label{bs-zeromode}
\end{equation}
Then, it has the correct ghost number $4$ 
and satisfies the level matching and
reality conditions. 
Note also that it has the nonvanishing component
only in the $\ket{j_\phi}$ sector in \eq{J}, 
which couples to the $\ket{\phi}$ sector 
as shown in \eq{source-component}.
Here, from \eq{QB=0} and the level matching condition, 
we can prove easily that $J$ is also BRST invariant:
\begin{equation}
 QJ = 0\,. 
\end{equation}
Comparing this to the equation \eq{conservation}, 
it is obvious that the boundary state $\ket{B_p}$ is 
a source of the closed string field only when 
the closed string coupling constant vanishes. 
In other words, when $g\ne 0$, 
the usual BRST invariant boundary state cannot be 
a source for the closed string field (unless $B_p*\Phi=0$). 
This means that, if the closed string coupling is turned on, 
the boundary state must be modified so that it satisfies 
the condition, 
\begin{equation}
 \delta_{\rm B} J = 0.
 \label{delj} 
\end{equation}
We can then regard this $J$ as a matter current 
that truly describes a D-brane. 
The necessity of this modification is not surprising. 
In fact, 
it is consistent with the usual picture of the string 
perturbation theory: 
Since a D-brane is a non-perturbative object and 
has mass $\sim 1/g$, 
it is infinitely heavy in the limit $g\rightarrow 0$ so that
it behaves as a rigid hyperplane in the space-time 
and this defines a conformal background. 
When small $g$ is turned on, it is still heavy but can receive some
recoil effect from the bulk and behaves as a non-relativistic object
moving in the space-time. To maintain the unitarity, there should be
collective coordinates for the D-brane \cite{collective} 
and they give the dynamical
degrees of freedom for the source.
Therefore, it is quite natural to assume that this modification 
is due to the open string excitation: it is schematically written as
\begin{equation}
 \ket{J} = e^{-S_b[X]}\ket{B_p}, 
 \label{bdry-int}
\end{equation}
where $S_b[X]$ is an appropriate boundary interaction \cite{CLNY}. 
In the presence of the boundary state $\ket{B_p}$ alone, 
the space-time symmetry 
such as the translational symmetry in the $x^i$ direction
is generally broken.
On the other hand, 
since the modified current $J$ contains the collective coordinate for
the broken symmetry, e.g., the scalar fields on the D-brane, 
it can keep the global gauge symmetry.
As a result, the current becomes a dynamical source and 
the equation \eq{delj} would effectively describes the behavior of 
the open string excitations on the D-brane. 
In the next subsection, we will discuss this point in more detail.

The same discussion can be applied to 
the rolling tachyon boundary state \cite{rolling}. 
The rolling tachyon is defined by inserting exact marginal 
tachyon vertices at the world-sheet boundary and is expected 
to be a solution which describes the rolling down of the open string
tachyon from the top to the bottom of the potential. 
According to the conjecture made by A.~Sen \cite{Sen}, 
it is believed that it describes the process that 
the unstable D-brane system decays 
by emitting closed strings \cite{LLM}. 
However, as for the usual boundary state, 
the rolling tachyon boundary state $\ket{B}_{\rm rolling}$ can be 
a source of a closed string field only when $g$ vanishes,%
\footnote{
Another possibility is  
that ${B}_{\rm rolling}$ and the classical
solution of the closed string field $\Phi$ satisfy 
the relation ${B_{\rm rolling}*\Phi}=0$. 
However, it is highly nontrivial. 
}
because it satisfies the condition, 
\begin{equation}
 Q\ket{B}_{\rm rolling} = 0. 
 \label{condition-rolling}
\end{equation} 
In fact, as pointed out in \cite{t-matter}
the (not covariant) conservation law  
of the energy momentum tensor of the D-brane,  
\begin{equation}
 \del^{\mu} T_{\mu\nu} = 0\,,
\end{equation}
follows from the condition \eq{condition-rolling}. 
This means that 
the energy exchange between D-brane and the bulk closed strings 
is completely ignored.%
\footnote{
See also the similar argument based on the low energy effective theory 
\cite{DJ}
and based on the toy model for open-closed SFT
\cite{ohmori,JV}.
}
Our claim is that
if the closed string coupling $g$ is turned on, 
we must modify not only the classical solution of $\Phi$ 
but also the source so that it satisfies the equation
\eq{conservation}. 
Here the modification would be again the form \eq{bdry-int} with
$\ket{B_p}$ is replaced with $\ket{B}_{\rm rolling}\,$.  
Note that if the original BRST invariant boundary state 
$\ket{B}$ is stable,
such as the BPS D-brane boundary state in Type II string
theory, the modified state $\ket{J}$ in \eq{bdry-int}
should still represent a D-brane. 
However, if it is unstable, such as bosonic D-branes 
or the rolling tachyon boundary state, 
then $\ket{J}$ need not express a D-brane but  
may decay into something by the condensation of the open string. 
Anyway, 
the classical solution of the bulk closed string field is deformed
by backreaction from the boundary state
and the boundary state is also deformed by the backreaction from the bulk 
so as to satisfy \eq{eom-source} and \eq{conservation} simultaneously.  
We note here that the total energy conservation 
is still guaranteed by \eq{total-conservation}. 
As a result, the obtained boundary state would 
correctly describe the decaying D-brane 
with emitting closed strings.


\subsection{Perturbative Expansions}
\label{sec:perturbative}

In this subsection, we sketch a method to solve the equations 
\eq{eom-source} and \eq{conservation} 
in which we start with a rigid boundary state satisfying 
$Q\ket{B}=0$ and then deform it perturbatively. 
It is closely related to the viewpoint of the usual world-sheet theory.

We first expand both the closed string field and the boundary state 
in the closed string coupling $g$ as 
\begin{equation}
 {\Phi} = \sum_{n=0}^{\infty}g^n{\Phi_n}, \qq
  {J} = \sum_{n=0}^{\infty}g^n {J_n}.
  \label{expansion}
\end{equation}
We take the lowest component as the (rigid) boundary state:
\begin{equation}
 \ket{J_0} = c_0^- \ket{B_p}\,. 
  \label{asuume}
\end{equation} 
Note that both the expansion begin with the zero-th order in $g$.
It is understood by the corresponding low energy theory 
(see Appendix \ref{freeSFT}). 

By substituting \eq{expansion} into the equations 
\eq{eom-source} and \eq{conservation}, 
we obtain the recursion formulae for $n+1\geq 0$,  
\begin{equation}
 \begin{cases}
  Q{\Phi_{n+1}} &= -{J_{n+1}} - 
   \displaystyle\sum_{m=0}^n
  {\Phi_m*\Phi_{n-m}}, \\
  Q{J_{n+1}} &= 
  2\displaystyle\sum_{m=0}^n {J_m*\Phi_{n-m}}. 
 \end{cases}
 \label{formula-pre}
\end{equation}
These equations say that, with given $J_0$, other components
$J_n\, (n\geq 1)$ and $\Phi_n\, (n\geq 0)$ will be  
determined recursively. 
To be precise, each component has two sectors 
according to the structure of the ghost zero-modes. 
Now we make the following ansatz to this ghost structure.
First, all the component $J_n\, (n\geq 1)$ 
is in the same sector as $J_0$, that is, in the sector $\ket{j_\phi}$.  
This indicates that only the physical sector 
$\ket{\phi}$ is coupled to it.
Correspondingly, we restrict the component fields of 
the string field to the physical target-space fields. 
That is, we require 
\begin{equation}
 \ket{\Phi} = c_0^-\ket{\phi}, 
 \label{restrict}
\end{equation} 
together with
\begin{equation}
 M^+ \ket{\phi} = 0.
  \label{extra}
\end{equation} 
The last condition is
necessary to eliminate such fields in $\ket{\phi}$ 
as do not couple to the boundary state. 
For example, the field $S(x)$ in the expansion \eq{component-phi} 
is eliminated by the condition \eq{extra}. 
The meaning of this will become clearer below. 

With these simplifications, 
the expansion \eq{expansion} is rewritten as 
\begin{equation}
 \ket{\Phi_n} \equiv c_0^- \ket{\phi_n},\qq
  \ket{J_n} \equiv c_0^-c_0^+ \ket{j_n}\,. 
\end{equation}
Then each equation in \eq{formula-pre} are decomposed into two sectors
as 
\begin{equation}
 \begin{cases}
  Q' \,\ket{\phi_{n+1}} &= 0\,, \\
  M^+ \,\ket{j_{n+1}} &= 0 \,,  
 \end{cases}
 \label{formula2}
\end{equation}
and 
\begin{equation}
 \begin{cases}
  L_0^+\,\ket{\phi_{n+1}} &= \ket{j_{n+1}} 
  + b_0^+b_0^-
  \displaystyle\sum_{m=0}^{n}\ket{\Phi_m*\Phi_{n-m}}\,, \\ 
  Q' \,\ket{j_{n+1}} &= 2b_0^+b_0^-
  \displaystyle\sum_{m=0}^{n}\ket{J_m*\Phi_{n-m}}\,. 
  \end{cases}
  \label{formula}
\end{equation}
Here, equations \eq{formula2} are in the sector $c_0^-\ket{\cdots}$,
whereas equations \eq{formula} are in the sector
$c_0^-c_0^+\ket{\cdots}$.  
We have used the assumption that both of ${\Phi*\Phi}$ and ${J*\Phi}$ 
are in the latter sector 
under the conditions \eq{asuume}, \eq{restrict} and \eq{extra}.%
\footnote{It is true for $\Phi *\Phi$ 
and we have ascertain it explicitly for massless sector of $J*\Phi$.
But we do not have concrete proof yet.
} 
We have eliminated the ghost zero-mode factor $c_0^-c_0^+$ 
by multiplying $b_0^+b_0^+$ in front of the $*$-products.

The equations in \eq{formula2} are constraint equations 
for the allowed degrees of freedom.
The first equation in \eq{formula2} together with \eq{extra} 
requires that 
the non-zero component state in the closed string field 
be in ``off-shell but physical'' states \cite{RT}. 
Recalling that $(Q')^{2}\propto M^+L_0^+$, 
the operator $Q'$ is nilpotent on the restricted space 
satisfying \eq{extra}.
Then first equation in \eq{formula2} says that $\ket{\phi}$ 
is in the $Q'$-cohomology. 
As seen from the definition of $Q'$ \eq{Qdef}, 
it means the physical state condition except for the on-shell
condition. 
Such a state is also called as the softly off-shell state.
Since the classical solution is in general 
off-shell (i.e., not the solution of the free equation of motion), it 
is a suitable condition. 
In fact, for the massless state, it gives the usual harmonic gauge
condition, after appropriate field redefinitions \cite{CLNY}.

On the other hand, 
the second equation of \eq{formula2} 
says that the source $\ket{j_n}$
still have one of the same property as the original boundary state
$\ket{B_p}$, i.e., $M^+\ket{B_p}=0$ in \eq{MB=0}. 
This guarantees that the fields which are coupled to 
the boundary state satisfies the condition \eq{extra} 
even after turning on the closed string coupling $g$. 

Under the constraints given by the equations \eq{formula2}, 
the equations \eq{formula} determine the classical solution 
of the string field and the consistent boundary state. 
To understand the structure of the equations \eq{formula}, 
it is useful to write down 
the first few equations in \eq{formula},  
\begin{align}
 \label{sln-0}
 Q' \, \ket{j_0} & = 0\,, \\
 \label{sln-1}
 L_0^+\,\ket{\phi_0} &= \ket{j_0}\,, \\
 \label{recoil-1}
 Q' \, \ket{j_1} &= 2b_0^+b_0^-\,\ket{J_0*\Phi_0}\,, \\
 \label{sln-2}
 L_0^+\, \ket{\phi_1} &= \ket{j_1} + b_0^+b_0^-\, 
 \ket{\Phi_0*\Phi_0} \,, \\
 &\ \ \vdots  \nonumber 
\end{align} 
From these equations, it is clear that, 
once the first component of the boundary state $\ket{j_0}$ 
is given, 
the equations \eq{formula} 
determine $\ket{\phi_n}$ and $\ket{j_n}$ successively. 
Below, we make some comments on each of the equations 
\eq{sln-0}--\eq{sln-2}:

The first equation
\eq{sln-0} is satisfied by our assumption \eq{asuume}. 
Since the action of $Q'$ on $\ket{j_0}$ is same as that on
$\ket{B_p}$, given by 
\begin{equation}
Q'\ket{B_p}=\sum_{n\ne 0}
 c_{-n}\left(L_n^{(m)}-\widetilde{L}_{-n}^{(m)}\right)\ket{B_p} \,, 
 \label{QQdef}
\end{equation}
it (with the level matching condition) states that the presence of the
source $\ket{j_0}$ keeps the conformal
invariance.
Together with the second equation in \eq{formula2}, it is equivalent
to the BRST invariance for the source in the lowest component 
(see \eq{MB=0}). 
Moreover, it implies that we can
start with any type of BRST invariant boundary states, which
is considered to be the conformal background.
For example, a boundary state with a constant electric or magnetic
flux turned on, that with traveling waves, the rolling tachyon
boundary state, and so on.

The second equation  
\eq{sln-1} 
carries the information on the off-shellness in the presence of the
source $\ket{j_0}$.
It is nothing but the the equation of motion in the case of
free SFT with a source term.
As discussed in Ref.\ \cite{DiVecchia-1}, 
it determines the leading term of the classical solution
(i.e., long range behavior)  
of the bulk fields when there is a BRST invariant boundary state. 
As originally discussed in \cite{CLNY}, it is also related to the
cancellation for divergences: 
in the cylinder diagram a closed string IR divergence comes from the
long cylinder limit, and it is canceled by
adding a disk diagram with a closed string insertion.
In other words, the presence of the boundary induces the closed 
string tadpole.

The third equation \eq{recoil-1} determines the first order 
modification 
of the boundary state from the original one, $\ket{j_0}$. 
It is necessary because of the breaking of the original 
conformal invariance by the closed string tadpole $\phi_0$.
Namely, 
it is the backreaction on the source coming from the change of the bulk.
As mentioned in the previous subsection, 
it is natural to interpret that the change is due to 
an open string excitation on the D-brane. 
Then, the equation \eq{recoil-1} 
says that the open string excitation is induced 
by the insertion of the closed string tadpole on the disk. 
It strongly suggests that when the tadpole getting closer to the
boundary, the operator product expansion of the closed string vertex 
with its mirror image causes a divergence and it is 
canceled by this open string vertex insertion. This is the similar
situation of the D-brane recoil \cite{collective, recoil} 
where the annulus
divergence coming from open string IR regime is canceled by the 
open string non-local insertion, 
although the correct relation between our analysis and these works 
are not yet clear.

The fourth equation \eq{sln-2} can determine the next leading term 
of the classical solution. 
The physical interpretation of this equation is obvious, 
that is, 
the first term of \eq{sln-2} means the backreaction coming 
from the change of boundary 
in the same way as \eq{sln-1} 
and the second term comes from the source due to
the self-interaction of the string field. 

In this way, 
once a BRST invariant source is given,  
both of the backreaction on the bulk and the boundary 
can be determined successively.

\resection{Conclusion and Discussion}

In this paper, we presented a general framework for the closed strings
in the presence of an arbitrary matter current. 
Starting with a closed string field theory 
with an arbitrary source term, 
we derived a couple of equations, 
one is the equation of motion for the closed string field, 
and the other is a constraint equation which expresses 
the covariant conservation law for the matter current. 
We discussed that we need both of the equations to describe 
the classical behavior of closed strings in the presence of 
the matter current. 
We also argued that our discussion can be regarded as 
a generalization of that of the general relativity, 
including the contribution from full massive fields. 
Then we applied our argument to D-branes. 
We claimed that the usual BRST invariant boundary state is not 
a consistent source, but it should be modified by turning on dynamical
degrees of freedom so that it satisfies the constraint equation. 
By perturbative expansion, 
we saw that the equation of motion and the covariant
conservation condition describe the backreaction on the source and
on the bulk, successively. 
This also suggests that the dynamical degrees of 
freedom are due to open string excitation.  

Since this is our first attempt to take into account 
the D-brane dynamics in the theory of off-shell closed strings,  
there are many issues that are not discussed in this paper 
and remained to be done. 
First, we should apply our method sketched in \S \ref{sec:perturbative} 
concretely to some definite matter, for example, a boundary state. 
From our discussion, it is expected that we would obtain 
a classical solution of the closed string field 
in the presence of D-brane with open string excitations. 
To perform it explicitly, 
we need the detail of the $*$-product. 
Technically, 
the calculation of the $*$-product can be done either by the
oscillator formalism or by the CFT technique.  
In the former case, it would be useful to use the level truncation 
approximation \cite{Taylor} even for the closed SFT.  
The approach using the CFT technique 
might also help to clarify the relation of our condition to the
usual world-sheet picture.


Another interesting issue is to apply our argument to 
a time dependent matter source. 
In our formalism, it is possible, at least formally, 
to obtain a solution which really describes the decaying process 
to the vacuum through the emission of the closed strings. 
Practically, we can use the rolling tachyon boundary state 
as the starting point and modify it by the perturbation as 
explained in \S \ref{sec:perturbative}. 
It is a fascinating issue to decide whether 
the modified boundary state starting from the rolling tachyon boundary 
state correctly describes the decaying process of non-BPS D-branes. 
Our argument could also apply to gravity theories. 
For example, if we find a solution that describes 
the decaying process of some extended object, 
the low energy limit might express the classical solution 
for the black hole evaporation. 
It may be also interesting to apply it to the D-brane inflation   
which is the original form of the inflationary brane model \cite{KK}. 
Furthermore, the system we proposed in this paper includes the 
self-interaction of string fields, so the low energy limit of it 
would have something to do with self-gravitating brane models \cite{SKOT}.  
Note, however, that
in order to relate the target-space field contained in the SFT to 
that of the low energy gravity, some field redefinition is needed.

Applying our argument to the superstring theories is also 
one of the important future works,  
although there are technical difficulties coming from the closed
super-SFT.  
If it is overcome, we could consider sources with NSNS or RR
charged objects and discuss various dualities. 
For instance, our setting seems quite useful 
to understand the AdS/CFT correspondence at the more fundamental level.  
The essence of the AdS/CFT correspondence is the duality 
between the open string theory on a D-brane 
and the closed string theory in the background of 
the classical geometry made by the D-brane. 
Recalling that our analysis produces both of 
the classical configuration of the closed string field 
and the open string excitation of the D-brane 
simultaneously, 
the AdS/CFT correspondence (more generally, the open/closed duality) 
might appear in the analysis. 
This would be worth considering even 
in the bosonic string field theory.

Finally, we make a comment of the relation between 
our formalism using a matter current 
and the quantum theory that governs the dynamics of the matter. 
In a realistic model, the matter current which we have considered 
through out this article is thought to consist of 
some ``matter fields'' and there should be an appropriate action 
which describes their dynamics. 
Especially, in the case of the D-brane, 
the matter field would be the open string field, 
and thus, we may consider that 
we must investigate a consistent open/closed string field 
theory in the presence of the D-brane. 
However, although constructing such a SFT is actually 
important future subject, 
it is sufficient to treat the matter as a current 
in determining the classical behavior of the closed string field. 
A similar situation is seen in the Maxwell's theory. 
In fact, we can determine the classical configuration of 
the electromagnetic fields in the presence of 
an electric current  
even if we do not know the fact that 
the current consists of 
the electrons which are governed by the QED.


\section*{Acknowledgments}
The authors would like to thank to 
T.\ Kugo, T.\ Nakatsu, H.\ Kajiura, T.\ Takayanagi, 
M.\ Fukuma, T.\ Sakai, J.\ Nishimura, T.\ Suyama, H.\ Fuji, 
and M.\ Sakagami for helpful discussions. 
The work of T.A. was supported in part by 
JSPS Research Fellowships for Young Scientists.

\appendix 

\resection{Definition of the $*$-product in HIKKO's SFT}
\label{star-product}

In this appendix, we briefly summarize the definition of 
the $*$-product of HIKKO's closed SFT. 
We note that 
the interaction vertex of any kind of closed SFT 
can be defined in the same way.

In defining the $*$-product, 
using the CFT language makes the discussion clear and elegant. 
We first define LPP's 3-point vertex following Ref.\ \cite{LPP}, 
which is determined uniquely if we give three conformal
mappings $\left\{h_r\,|\,r=1,2,3\right\}$ from unit disks (with
coordinate $w_r$) to a
sphere (with coordinate $z$) as, 
\begin{equation}
 \bra{v_{123}^{\rm LPP}}\ket{\phi_3}_3\ket{\phi_2}_2\ket{\phi_1}_1
  \equiv \biggl\langle 
  h_3\left[\phi_3\right](z_3)\, 
  h_2\left[\phi_2\right](z_2)\, 
  h_1\left[\phi_1\right](z_1)
  \biggr\rangle_{S^2}, 
\end{equation}
where $\ket{\phi_r}_r\equiv\phi_r(w_r=0)\ket{0}_r\,(r=1,2,3)$ are 
arbitrary states on disks. 
The r.h.s. is the three point correlation function on $S^2$.  
Each map $h_r$ determines the conformal transformation 
of the vertex operator $\phi_r(w_r=0)$ at the origin on the disk 
to the one at $z_r$ on the sphere. 
For HIKKO's closed SFT, the conformal mappings 
$\left\{h_r\,|\, r=1,2,3 \right\}$ are defined as 
the composition of two conformal maps, 
$h_r\equiv f_{\rm M}\circ g_r: w_r\mapsto z$. 
Here, $g_r: w_r \mapsto \rho \,,$ is the mapping from disk $r$ to the
cylinder ({\em $\rho$-plane})%
\footnote{
Here, $w_r$ is assumed to be represented as 
$
 w_r = \exp\left(\tau_r+i\sigma_r\right) \, 
 (-\infty<\tau_r\le 0,\ 0\le\sigma_r< 2\pi). 
$
}
with 
\begin{equation}
 \label{map1}
 \rho = 
\begin{cases}
 \alpha_1 \ln w_1\,, \\
 \alpha_2 \ln w_2 + 2\pi i \alpha_1\,, \\
 \alpha_3 \ln w_3 + 2\pi i \left(\alpha_1+\alpha_2\right)\,, 
\end{cases}
\end{equation} 
and $f_{\rm M}:\rho \mapsto z \,, $ is given by (the inverse of) 
the Mandelstam mapping, 
\begin{equation}
 \rho = \sum_{r=1}^3 \alpha_r \ln \left(z-z_r\right). 
  \label{Mandelstam}
\end{equation}
The parameters $\left\{\alpha_r\,|\,r=1,2,3\right\}$ 
in \eq{map1} and \eq{Mandelstam} 
are the {\em string length parameters} \cite{hikko} which 
satisfy the condition, 
$\alpha_1 + \alpha_2 + \alpha_3 = 0$. 
Above map \eq{map1} corresponds to the case where 
$\alpha_1,\alpha_2 > 0,\,\alpha_3 < 0$.
Note that this construction is generalized to
arbitrary $N$-point vertex while we need only the $3$-point one here.
Using the LPP vertex above, 
the 3-point vertex of HIKKO's closed SFT is given by 
\begin{equation}
 \bra{V_{123}}\equiv 
 \int \, \prod_{r=1}^{3}\frac{d\sigma_r}{2\pi}\, d\alpha_r \,
 \delta(\alpha_1 + \alpha_2 + \alpha_3)\,
 \bra{v_{123}^{\rm LPP}}
  b_0^{(1)-}b_0^{(2)-}b_0^{(3)-}.
\end{equation}
Now the $*$-product for two string fields is defined by 
\begin{equation}
 b_0^{(4)-}\ket{\Phi*\Psi}_4 \equiv 
  \bra{V_{123}}\ket{R(3,4)}\ket{\Phi}_2\ket{\Psi}_1, 
\end{equation}
or equivalently, combining with the inner product as
\begin{equation}
 \Lambda\cdot(\Phi*\Psi) \equiv 
  \bra{V_{123}}\ket{\Lambda}_3\ket{\Phi}_2\ket{\Psi}_1.
\end{equation}

\resection{Low Energy Effective Action of the Free Closed String Field
Theory}
\label{freeSFT}

In this appendix, we explicitly decompose the closed string field 
into component fields and show that the low energy action of 
the free part of the SFT action \eq{kinetic} 
reproduces the quadratic part of 
the low energy effective action of the bosonic string theory. 
After that, we write down the gauge transformation of 
the component fields explicitly and show that the $g\to0$ limit 
of the gauge transformation \eq{gauge-trans} gives 
the proper gauge transformations of the gravity fields. 

We decompose $\ket{\phi}$ and $\ket{\psi}$ as following, 
\begin{align}
 \label{component-phi}
 \ket{\phi} &= \int {d^{26}k \over (2\pi)^{26}}
  \biggl\{\biggl[
    {T}(k) 
  + {1\over 2\sqrt{2}}
    \widehat{h}_{\mu\nu}(k)
    \left(
    \alpha_{-1}^{\mu}\talpha_{-1}^{\nu}+\talpha_{-1}^{\mu}\alpha_{-1}^{\nu} 
    \right) \nn
  &\qq\qq\qq\quad + {1\over 2\sqrt{2}} 
    {B}_{\mu\nu}(k)\left(
    \alpha_{-1}^{\mu}\talpha_{-1}^{\nu}-\talpha_{-1}^{\mu}\alpha_{-1}^{\nu}
    \right) \nn
  &\qq\qq\qq\quad - {1\over \sqrt{2}}
    \widehat{D}(k)\left(c_{-1}\tb_{-1}+\tc_{-1}b_{-1}\right) \nn 
  &\qq\qq\qq\quad + {1\over \sqrt{2}}
    {S}(k)\left(c_{-1}\tb_{-1}-\tc_{-1}b_{-1}\right)
  + \cdots 
 \biggr]c_1\tc_1\ket{k}\biggr\} \,,  
\end{align}
\begin{align}
 \ket{\psi}&= \int {d^{26}k \over (2\pi)^{26}}
   \biggl\{\biggl[
    -{i\over \sqrt{2}}b_{\mu}(k)\,
  \left(b_{-1}\talpha_{-1}^{\mu}+\tb_{-1}\alpha_{-1}^{\mu}\right) \nn
  &\qq\qq\qq\quad +{i\over \sqrt{2}}e_{\mu}(k)\,
  \left(b_{-1}\talpha_{-1}^{\mu}-\tb_{-1}\alpha_{-1}^{\mu}\right)
  +\cdots \biggr]c_1\tc_1\ket{k} \biggr\} \,. 
  \label{component-psi}
\end{align}
From the reality condition \eq{reality}, 
we see that all the component fields are real; $T^{\ast}(k)=T(-k)$. 
Substituting the expansion \eq{component-phi} and 
\eq{component-psi} into the free SFT action \eq{kinetic}, 
we obtain 
\begin{align}
 S_0 =\int d^{26}x \biggl\{
 &-{1\over2}T\left(\del^2-2\right)T 
 +{1\over4}\widehat{h}^{\mu\nu}\del^2\widehat{h}_{\mu\nu} 
 -{1\over4}{B}^{\mu\nu}\del^2{B}_{\mu\nu}
 +{1\over2}\hD\del^2\hD 
 -{1\over2}S\del^2 S \nn 
 &-b_{\mu}
   \left(\del^{\nu}\hh_{\mu\nu}+\del_{\mu}{\hD}\right)
  -e_{\mu}
   \left(\del^{\nu}{B}_{\mu\nu}-\del_{\mu}{S}\right) 
 +{1\over2}b^2 +{1\over2}e^2
 \biggr\} \,. 
 \label{grav-1}
\end{align}
In this action, $b_{\mu}(x)$ and $e_{\mu}(x)$ are 
auxiliary fields and can be integrated out. 
At the same time, we redefine the fields 
$\hh_{\mu\nu}$ and $\hD$ as 
\begin{equation}
 \hh_{\mu\nu}\equiv h_{\mu\nu}+\eta_{\mu\nu}D,\qq
  \hD \equiv D + {1\over2}h^{\mu}_{\ \mu}\,.  
\end{equation}
The obtained result is%
\footnote{
We have used the fact that the field $S(x)$ can be gauged away, 
which we will mention in the end of this appendix. 
}
\begin{align}
 S_0 &= \int d^{26}x \biggl\{
   {1\over2}\left[\left|\del T\right|^2+2T^2\right]  \nn
   &\qq\qq\quad 
    -\frac{1}{2\kappa^2}\left(\sqrt{-g}R\right)_2 
    + 6\left|\del D\right|^2 
    +{1\over12} \left|H_{\mu\nu\rho}\right|^2  
	      \biggr\}\,,
 \label{quadratic}
\end{align}
where 
the metric is expanded around $\eta_{\mu\nu}$ as 
$g_{\mu\nu}\equiv \eta_{\mu\nu}+\kappa h_{\mu\nu}$ and  
\begin{equation} 
 -\frac{1}{2\kappa^2}\left(\sqrt{-g}R\right)_2 
 \equiv  
 -{1\over4}h^{\mu\nu}\left(
      \del^2h_{\mu\nu}-2\del_{\nu}\del^{\rho}h_{\mu\rho}
      +2\del_{\mu}\del_{\nu}h^\rho_{\ \rho}
      -\eta_{\mu\nu}\del^2 h^\rho_{\ \rho}\right)\,,
\end{equation}
is the quadratic part of the Einstein-Hilbert Lagrangian.
$H_{\mu\nu\rho}$ represents the field strength of $B_{\mu\nu}$, 
\begin{equation}
 H_{\mu\nu\rho}\equiv 
  \del_{\mu}B_{\nu\rho} 
  +\del_{\nu}B_{\rho\mu} 
  +\del_{\rho}B_{\mu\nu}\,. 
\end{equation} 
Then we have shown that \eq{quadratic} reproduces  
the quadratic part of the low energy effective action 
of the bosonic string theory in the Einstein frame, and thus, 
it turns out that the component fields 
$T(x)$, $h_{\mu\nu}(x)$, $D(x)$ and $B_{\mu\nu}(x)$ 
actually correspond to the tachyon, graviton, dilaton and 
antisymmetric two tensor field, respectively. 
We note here that the expansion of the low energy effective action 
starts from the zero-th order in the gravitational coupling $\kappa$. 
Similarly, since the source term for the gravity theory is 
written as 
\begin{equation}
 S_0 \sim {1\over\kappa} \int d^{26}x 
  \Bigl[g^{\mu\nu}(x)T_{\mu\nu}(x)+\cdots\Bigr]\,, 
\end{equation}
the expansion of the source action also starts from 
the zero-th order in $\kappa$. 
Recalling $\kappa\propto g$, this fact guarantees the correctness 
for the perturbative 
expansion of $\Phi$ and $J$ in \eq{expansion}.

Next, we write down the gauge transformation of the component fields. 
To this end, 
we expand the gauge parameter string field 
$\ket{\Lambda} = c_0^-\ket{\lambda_1}+c_0^-c_0^+\ket{\lambda_2}$ 
as  
\begin{align}
 \ket{\lambda_1} &= \int \frac{d^{26}k}{(2\pi)^{26}}\Bigl\{
   \frac{i}{\sqrt{2}}\epsilon_{\mu}(k)
     \left(\talpha_{-1}^{\mu}c_1-\alpha_{-1}^{\mu}\tc_1\right) \nn
   &\qq\qq\qq
   -\frac{i}{\sqrt{2}}\zeta_{\mu}(k)
   \left(\talpha_{-1}^{\mu}c_1+\alpha_{-1}^{\mu}\tc_1\right)
   +\cdots \Bigr\}\ket{k}\,, \\ 
 \ket{\lambda_2} &= \int \frac{d^{26}k}{(2\pi)^{26}}\Bigl\{
   -\frac{1}{\sqrt{2}}\eta(k)
   +\cdots \Bigr\}\ket{k}\,, 
\end{align}
then the gauge transformations of the component fields in 
\eq{component-phi} and \eq{component-psi} become
\begin{alignat}{2}
 \delta {T} &= 0\,, & \qq \qq  & \nn 
 \delta \widehat{h}_{\mu\nu} 
   &= \del_{\mu}\epsilon_{\nu} + \del_{\nu}\epsilon_{\mu}\,, & 
 \delta {B}_{\mu\nu}
   &= \del_{\mu}\zeta_{\nu} + \del_{\nu}\zeta_{\mu}\,, \\
 \delta \widehat{D} &= \del \cdot \epsilon\,, &
 \delta {S} &= -\del \cdot \zeta + \eta\,, \nn
 \delta b_{\mu} &= \del^2 \epsilon_{\mu}\,, &
 \delta e_{\mu} &= -\del^2 \zeta_{\mu}+\del_\mu\eta\,. \nonumber
\end{alignat} 
In terms of the redefined fields $h_{\mu\nu}$ and $D$, 
the transformation becomes 
\begin{equation}
 \delta h_{\mu\nu}=\del_{\mu}\epsilon_{\nu}+\del_{\nu}\epsilon_{\mu}\,, 
  \qq
  \delta D = 0\,. 
\end{equation} 
From these relations, we see that fields in the low energy action 
\eq{quadratic} are surely to be the fields in the gravity theory. 
Moreover, as we mentioned in the section 2, 
the field $S(x)$ can be actually gauged away 
using the degree of freedom of the gauge parameter $\eta$.



\end{document}